\documentclass[%
preprint,
 showpacs,showkeys,
 amsmath,amssymb,
 aps,
]{revtex4-1}
\usepackage{amsmath}
\usepackage{cases}
\usepackage{amssymb}
\usepackage{CJK}
\usepackage{graphicx}
\usepackage{dcolumn}
\usepackage{bm}
\usepackage{color}
\usepackage[pagewise]{lineno}

\begin{document}
\begin{CJK*}{GBK}{} 

\preprint{APS/123-QED}

\title{A Scaling Phenomenon in Shannon Information Uncertainty Difference of fragments in Heavy-ion Collisions}

\author{Chun-Wang MA$^{1,2}$}\thanks{Email: machunwang@126.com}
\author{Yi-Dan SONG$^{2}$}
\author{Chun-Yuan QIAO$^{1}$}
\author{Shan-Shan WANG$^{1}$}
\author{Hui-Ling WEI$^{2}$}
\author{Yu-Gang MA$^{3}$}\thanks{Email: ygma@sinap.ac.cn}
\author{Xi-Guang Cao$^{3}$}

\affiliation{$^{1}$ Institute of Particle and Nuclear Physics, Henan Normal University, \textit{Xinxiang 453007}, China\\
$^{2}$ College of Physics and Electronic Engineering, Henan Normal University, \textit{Xinxiang 453007}, China\\
$^{3}$ Shanghai Institute of Applied Physics, Chinese Academy of Sciences, \textit{Shanghai 201800}, China\\
}




\date{\today}

\begin{abstract}
The Shannon information-entropy uncertainty (in brief as "information uncertainty") is used to analyze the fragments in the measured 140$A$ MeV $^{40, 48}$Ca + $^{9}$Be and $^{58, 64}$Ni + $^{9}$Be reactions. A scaling phenomenon is found in the information-uncertainty difference of fragments between the reactions. The scaling phenomenon is explained in a manner of canonical ensemble theory, and is reproduced in the simulated reactions by the antisymmetric molecular dynamics (AMD) and AMD + GEMINI models. The probes based on information uncertainty, requiring no equilibrium state of reaction, can be used in the non-equilibrium system, and bridge the results of the static thermodynamics models and the evolving dynamical transport models.
\end{abstract}
\maketitle
\end{CJK*}


\section{introduction}
The Shannon information entropy theory, which aims to measure the uncertainty in a random variable qualifying the expected value of the information contained in a message \cite{Shannon}, provides a constructive criterion for setting up probability distributions on the basis of partial knowledge \cite{Jaynes}. Besides its various applications, Shannon information entropy is introduced in the study of hadron collisions and heavy-ion collisions (HICs), which is very helpful for measuring the chaoticity in the hadron decaying branching process \cite{EntrHC} and probing the liquid-gas phase transition in nuclear disassembly \cite{YGMaZipfPRL99}. Some works dealing with the information entropy in fragmenting systems can also be found in Res. \cite{InfEnt02,InfEnt04}. With a slightly difference to the definition of Shannon information entropy for a reaction system, the Shannon information-entropy uncertainty ($In$,  in brief as "information uncertainty") is introduced to study the information carried by a specific fragment in HICs \cite{MaPLB15}. The system of HICs experiences the compressing and the expanding stages, in which nuclear matter changes from high density to low density and the properties of the nuclear matter are difficult to be probed. Meanwhile, the probes for nuclear matter in HICs basing on fragments are mostly model dependent. The measurable fragments in experiments, which carry partial information of system, can be used to investigate the characteristics of nuclear matter in an evolving nuclear reaction. Thus it is interesting to learn that the information uncertainty probes are independent of models, which helps the theoretical and experimental analysis \cite{MaPLB15,MaCTP14}. In this work, a scaling phenomenon in the information uncertainty of fragments will be presented.

\section{model description}

\begin{figure}[htbp]
\includegraphics
[width=8.6cm]{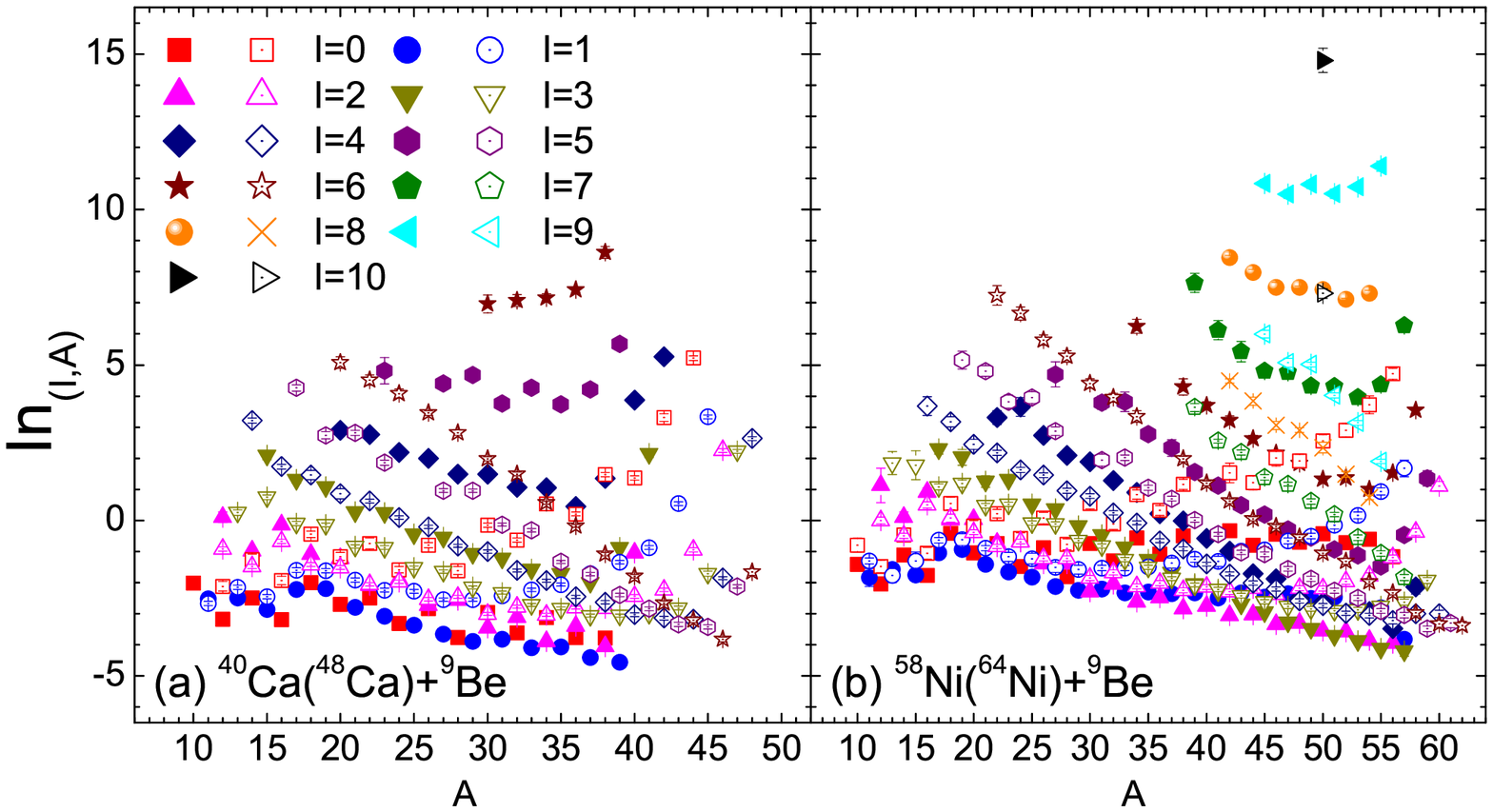}
\caption{\label{EntFrag} (Color online) Information uncertainty $In_{(I,A)}$ of fragment in the measured 140$A$ MeV $^{40, 48}$Ca + $^{9}$Be [in (a)] and $^{58, 64}$Ni + $^{9}$Be [in (b)] reactions. The results for the symmetric $^{40}$Ca ($^{58}$Ni) reaction and the neutron-rich $^{48}$Ca ($^{64}$Ni) reaction are denoted by full and open symbols, respectively.
}\label{FInIA}
\includegraphics
[width=8.6cm]{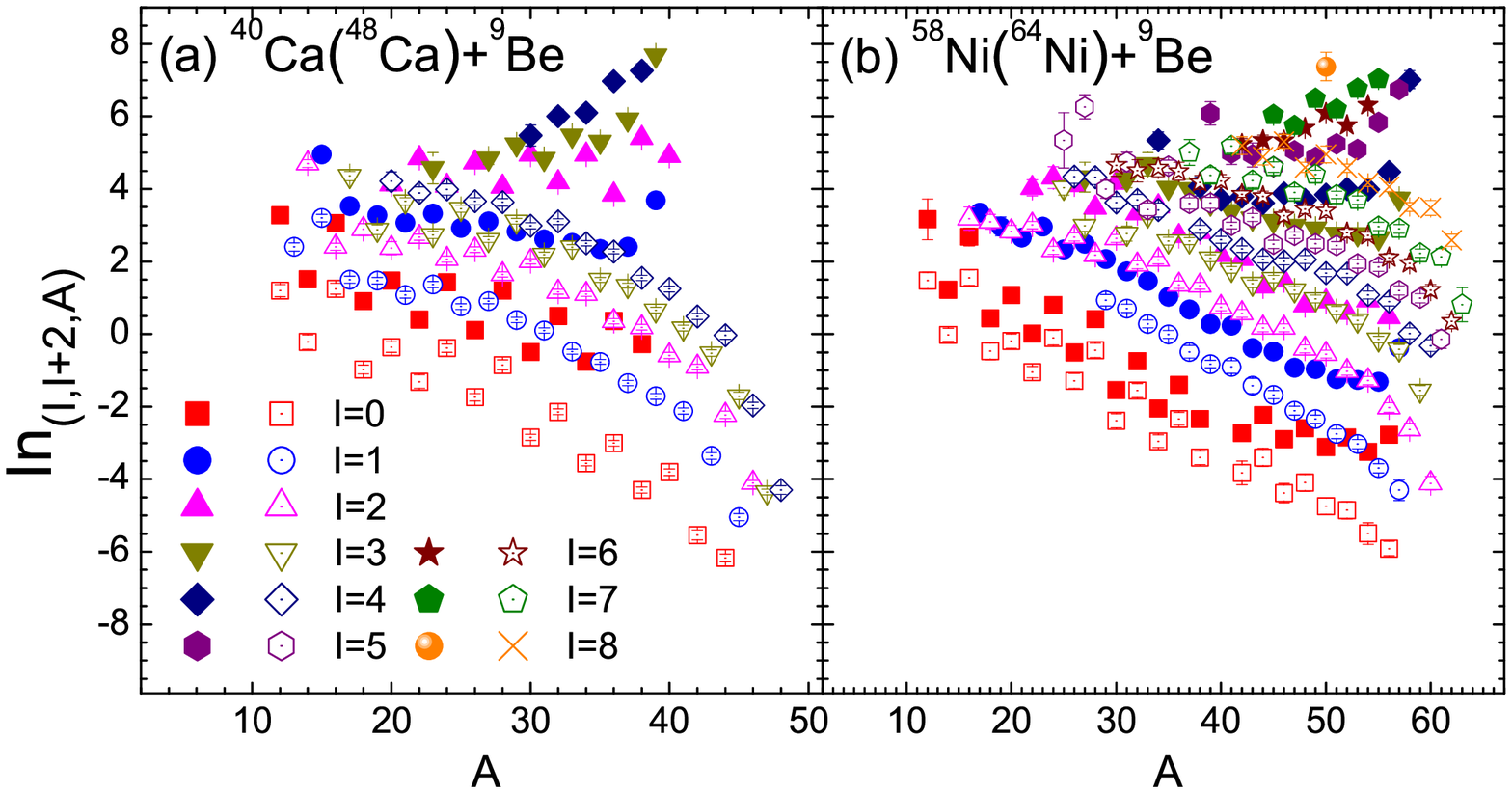}
\caption{\label{EnIsobDif} (Color online) Difference of information uncertainty between the (I,A) and (I+2,A) isobars [$In_{(I, I + 2, A)}$] in the 140$A$ MeV $^{40, 48}$Ca + $^{9}$Be [in (a)] and $^{58, 64}$Ni + $^{9}$Be [in (b)] reactions. The results for the symmetric $^{40}$Ca ($^{58}$Ni) reaction, and the neutron-rich $^{48}$Ca ($^{64}$Ni) reaction are denoted by full and open symbols, respectively.
}\label{FInI2A}
\end{figure}

The $In$ of an event in a system is related to its possibility. In HICs, fragments can be seen as the independent events in the reaction system, with the possibilities denoted by their cross sections. Following the definition in Ref. \cite{MaPLB15}, the $In$ of a fragment $(I, A)$ is written as,
\begin{equation}\label{Entp-FrgAI}
In_{(I, A)} = -\mbox{ln}\sigma(I, A),
\end{equation}
with $I \equiv N - Z$ being the neutron-excess of the fragment and $\sigma(I, A)$ being the cross section. In one reaction, the difference between the  $In$ of isobars differing 2 units in $I$ is,
\begin{equation}\label{Entp-FrgAI2}
In_{(I, I + 2, A)} = In_{(I, A)} - In_{(I + 2, A)}.
\end{equation}
Though $In_{(I, A)}$ defined in Eq. (\ref{Entp-FrgAI}) is not exact according to the Shannon information entropy theory, it is proven that $In_{(I, I + 2, A)}$ is correct because of the cancellation of system dependent parameters in $In$ \cite{MaPLB15}. In two reactions, the difference of $In_{(I, I + 2, A)}$ can be defined as,
\begin{equation}\label{Entp-FrgAI2R2}
\Delta_{21}In_{(I, I + 2, A)} = In_{2(I, I + 2, A)} - In_{1(I, I + 2, A)},
\end{equation}
with indexes 1 and 2 denoting the reactions. The results of $\Delta_{21}In_{(I, I + 2, A)}$ for the 140$A$ MeV $^{40, 48}$Ca + $^{9}$Be and $^{58, 64}$Ni + $^{9}$Be reactions, which have been measured by Mocko \textit{et al} at the National Superconducting Cyclotron Laboratory (NSCL) in Michigan State University \cite{Mocko06}, have been reported in Ref. \cite{MaPLB15}. In this letter, we will study the difference of information uncertainty of the fragment with large neutron-excess. In Fig. \ref{EntFrag}, the $In$ of the measured fragment is plotted according to $I$. It is found that $In_{(A, I)}$ increases with $I$ in the reactions, and $In_{(A, I)}$ also depends on $A$. In addition, $In_{(I, A)}$ depends on the asymmetry of reaction system.

\section{results and discussion}

In Fig. \ref{EnIsobDif}, $In_{(I, I + 2, A)}$ between the isobars are plotted. In the $^{40}$Ca reaction, from $I = $ 0 to 4, the trends of $In_{(I, I + 2, A)}$ change from decreasing to increasing with the increasing $A$, and $In_{(I, I + 2, A)}$ for the $I = $ 2 fragment chain becomes relative consistent. Similar changes are observed in $In_{(I, I + 2, A)}$ for the $^{58}$Ni reaction. $In_{(I, I + 2, A)}$ decreases with the increasing $A$ in the neutron-rich $^{48}$Ca and $^{64}$Ni reactions.

\begin{figure}[htbp]
\includegraphics
[width=8.6cm]{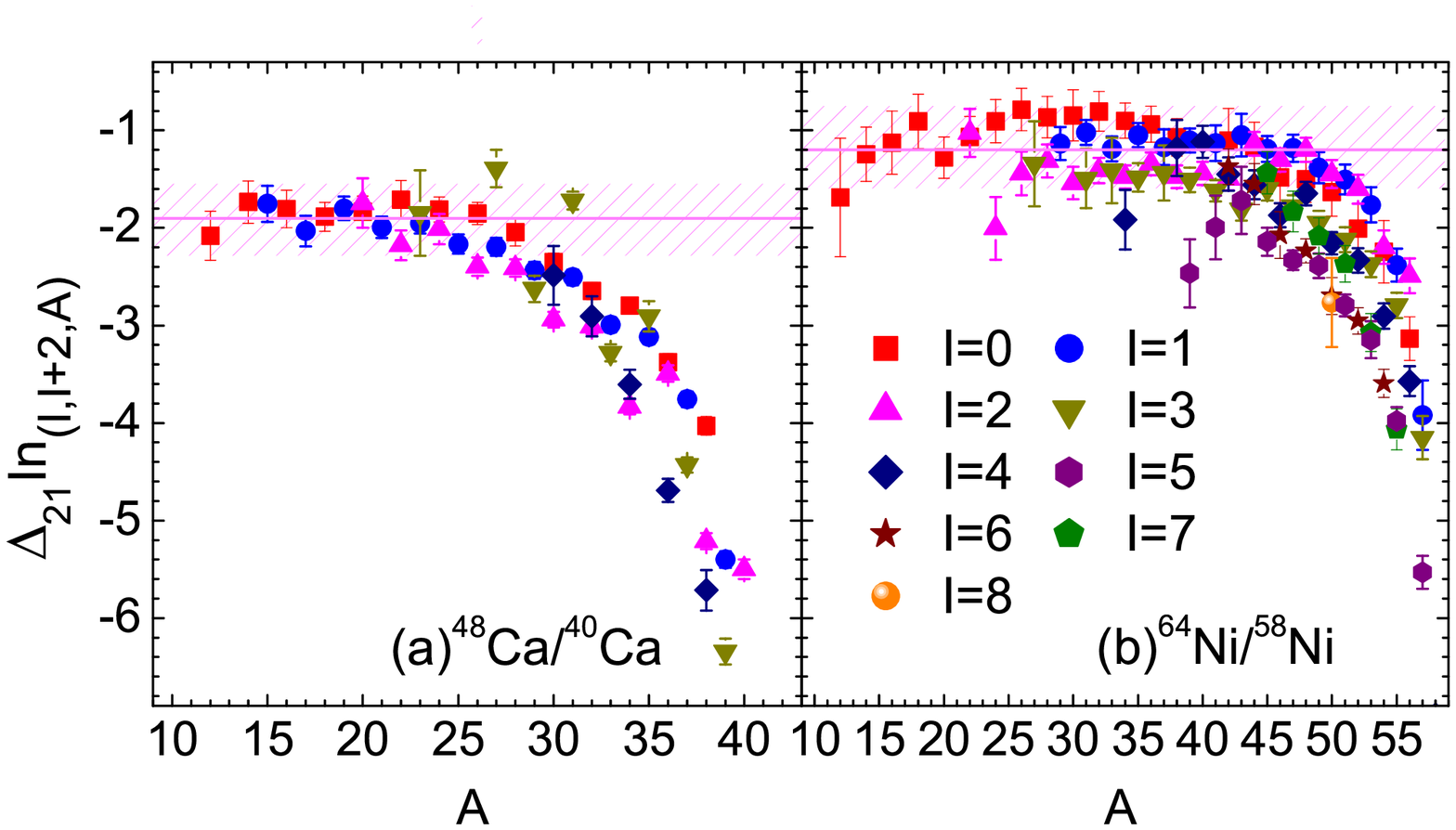}
\caption{\label{Endif2} (Color online) $\Delta_{21}In_{(I, I + 2, A)} = In_{2(I, I + 2, A)} - In_{1(I, I + 2, A)}$] between: in (a) the neutron-rich $^{48}$Ca and the symmetric $^{40}$Ca reactions, and in (b) the $^{64}$Ni and $^{58}$Ca reactions.
}\label{DFInI2A}
\includegraphics
[width=8.6cm]{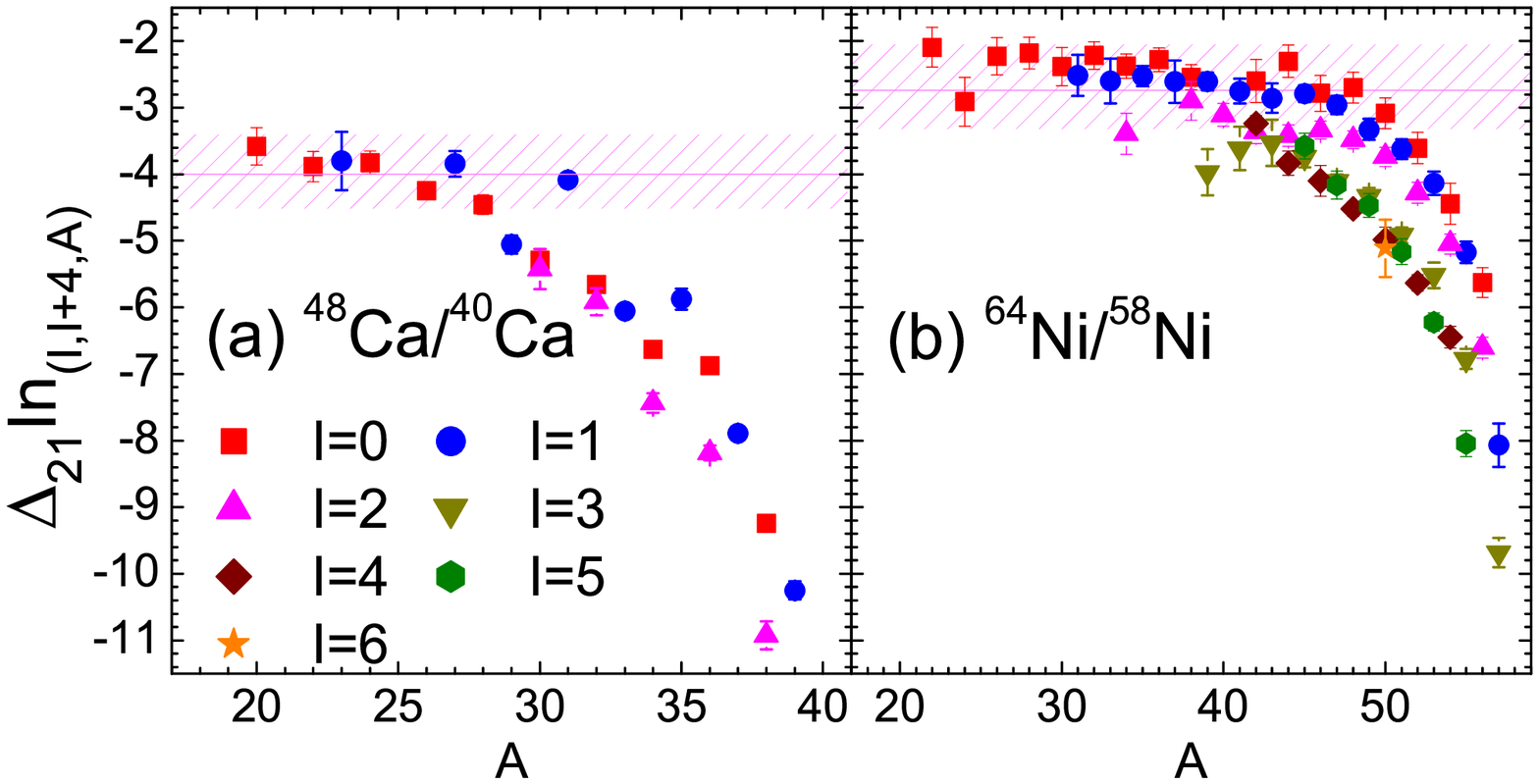}
\caption{\label{Endif4} (Color online) The same as Fig. \ref{Endif2} but for $\Delta_{21}In_{(I, I + 4, A)}$.
}\label{DFInI4A}
\includegraphics
[width=8.6cm]{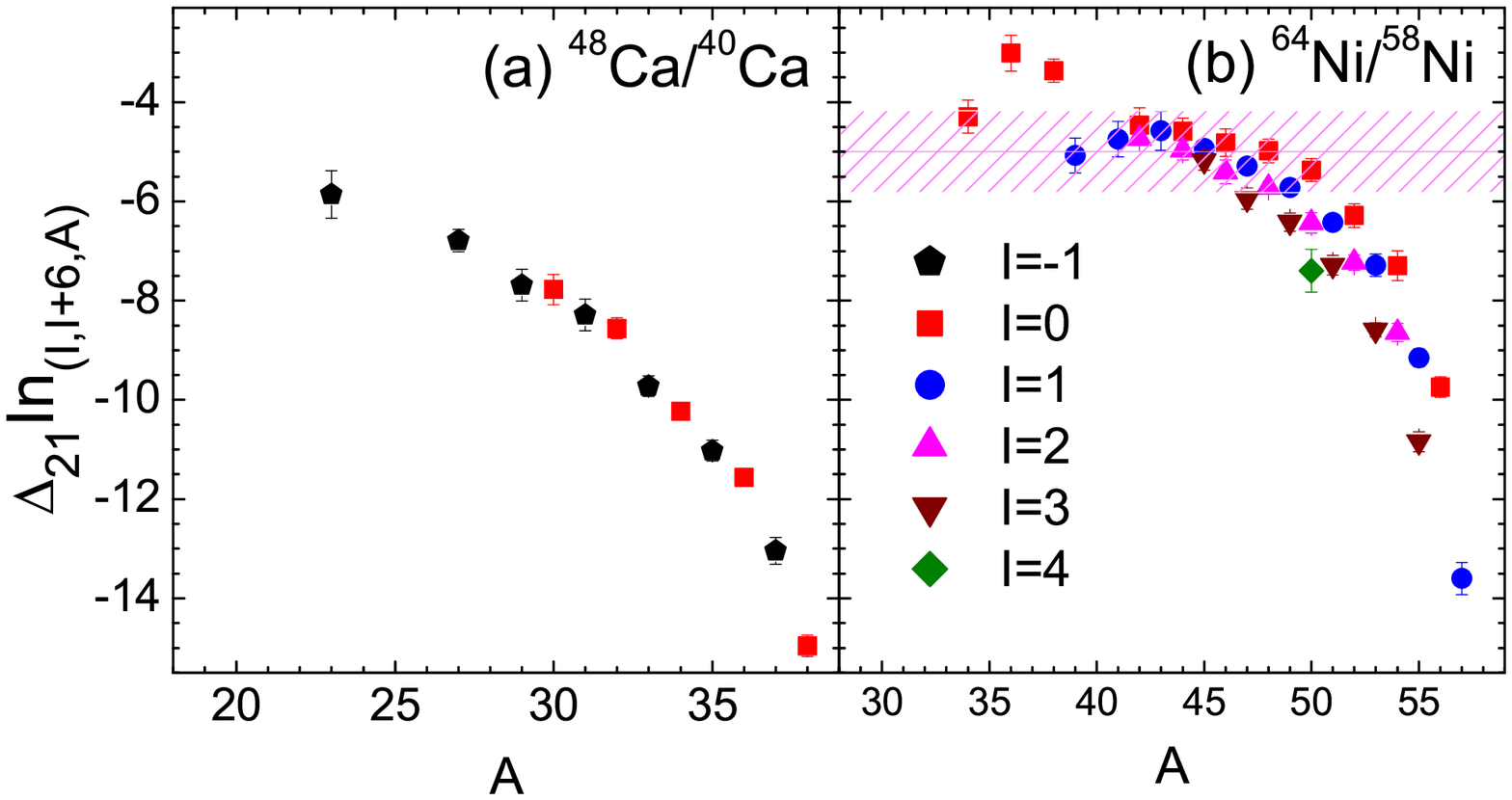}
\caption{\label{Endif6} (Color online) The same as Fig. \ref{Endif2} but for $\Delta_{21}In_{(I, I + 6, A)}$. 
}\label{DFInI6A}
\end{figure}

In Fig. \ref{Endif2}, $\Delta_{21}In_{(I, I + 2, A)}$ between the $^{48}$Ca and $^{40}$Ca (for simple, denoted as $^{48}$Ca/$^{40}$Ca) reactions, and that between the $^{64}$Ni and $^{58}$Ni  (denoted as $^{64}$Ni/$^{58}$Ni) reactions  (the index 1 for the neutron-rich system and 2 for the symmetric system) are plotted in panels (a) and (b), respectively. The shadowed areas in the figure illustrate the ranges in which $\Delta_{21}In_{(I, I + 2, A)}$ forms plateau. The plateaus have the values of -1.9 $\pm$ 0.3 and -1.2 $\pm$ 0.4 in (a) and (b), respectively. The $\Delta_{21}In_{(I, I + 2, A)}$ for  the $^{48}$Ca/$^{40}$Ca reactions are much more consistent than those for the $^{64}$Ni/$^{58}$Ni reactions. Some similar results have been shown in Ref. \cite{MaPLB15}.

For fragments with larger neutron-excess, similarly to the definition in Eq. (\ref{Entp-FrgAI2}), the difference of $In$ between isobars differing 4 units in $I$, $In_{(I, I + 4, A)}$, can be defined as,
\begin{equation}\label{Entp-FrgAI4}
In_{(I, I + 4, A)} = In_{(I, A)} - In_{(I + 4, A)},
\end{equation}
and $In_{(I, I + 6, A)}$ is defined as,
\begin{equation}\label{Entp-FrgAI6}
In_{(I, I + 6, A)} = In_{(I, A)} - In_{(I + 6, A)}.
\end{equation}
From Eq. (\ref{Entp-FrgAI4}), one can define the difference of $In_{(I, I + 4, A)}$ between two reactions as,
\begin{equation}\label{Entp-FrgAI4R2}
\Delta_{21}In_{(I, I + 4, A)} = In_{2(I, I + 4, A)}-In_{1(I, I + 4, A)}.
\end{equation}
And from Eq. (\ref{Entp-FrgAI6}), one can define the difference of $In_{(I, I + 6, A)}$ between two reactions as,
\begin{equation}\label{Entp-FrgAI6R2}
\Delta_{21}In_{(I, I + 6, A)} = In_{2(I, I + 6, A)}-In_{1(I, I + 6, A)}.
\end{equation}

\begin{figure}[htbp]
\includegraphics
[width=8.6cm]{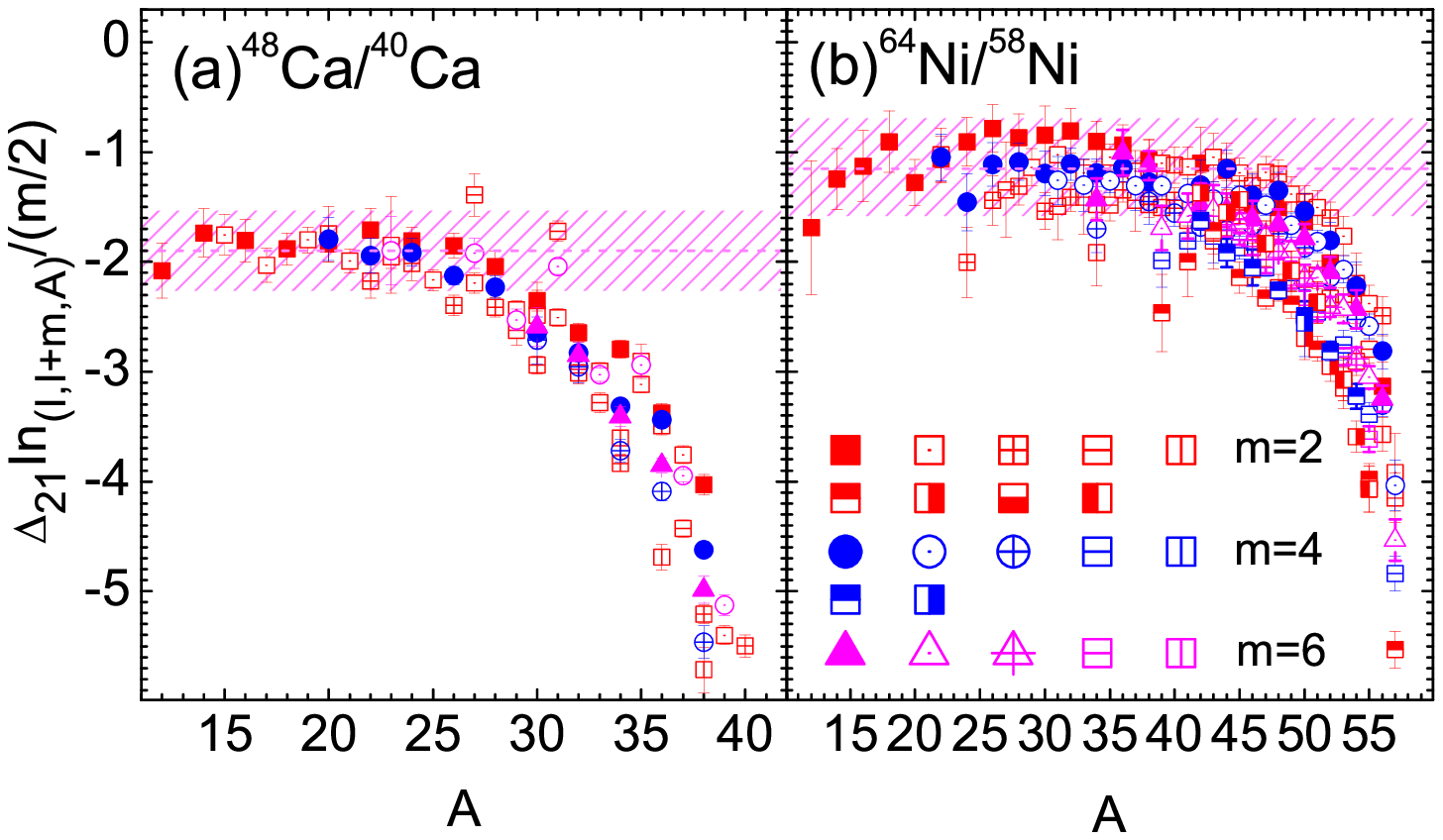}
\caption{(Color online) $S_{\Delta_{21}In}$  for the $^{48}$Ca/$^{40}$Ca reactions [in (a)], and for the $^{64}$Ni/$^{58}$Ni reactions [in (b)], with $m =$ 2, 4, and 6. The different filling of symbols denote the fragment with different $I$. The shadowed areas are the same as those in Fig. \ref{DDScaled}(b).
}\label{DDScaled}
\includegraphics
[width=8.6cm]{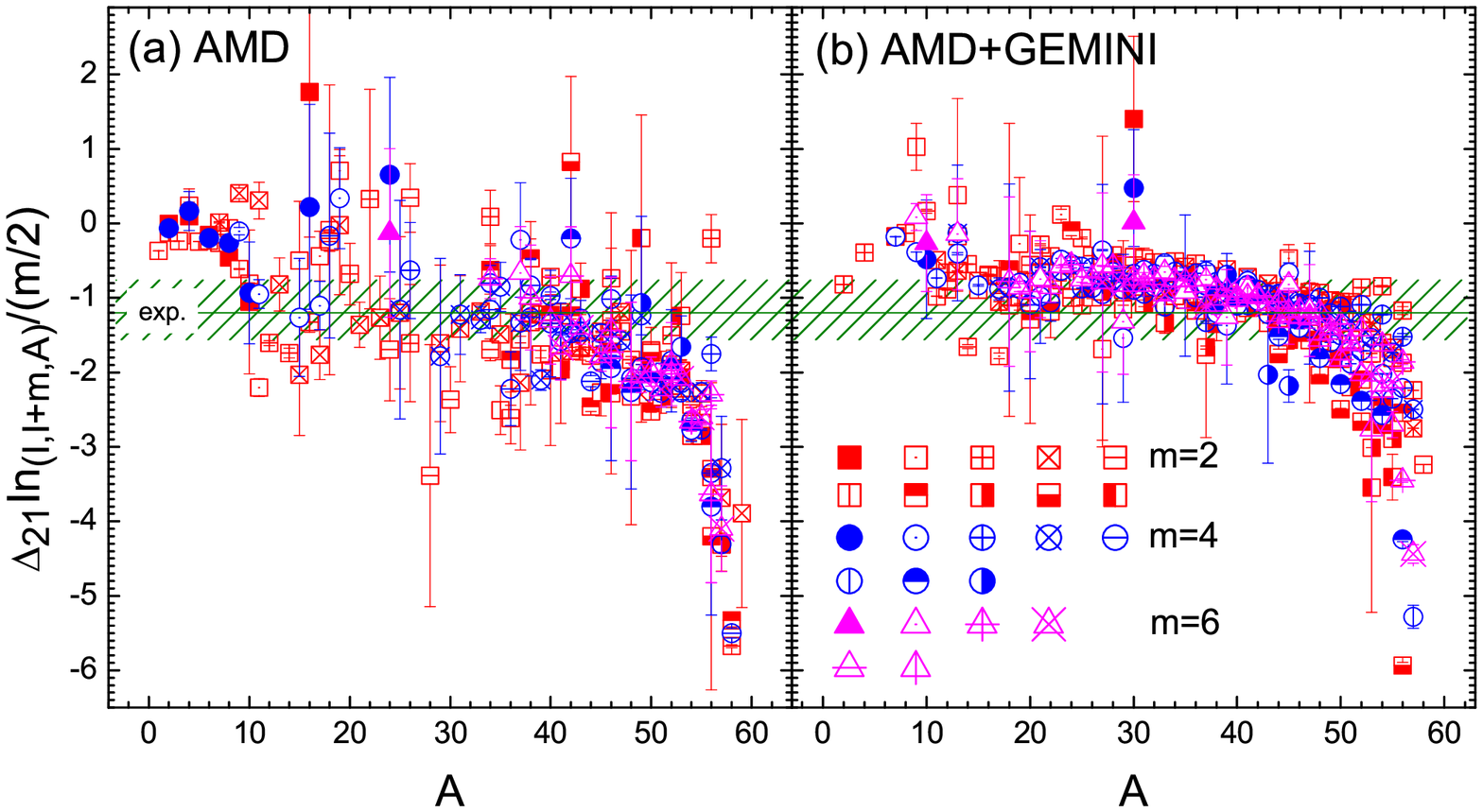}
\caption{(Color online) The $S_{\Delta_{21}In}$ for the $^{64}$Ni/$^{58}$Ni reactions calculated by AMD [in (a)], and by AMD + GEMINI [in (b)]. The shadowed area is experimental plateau range, which is the same as that plotted in Fig. \ref{DDScaled}(b).
}\label{DDScaledAMD}
\end{figure}

$\Delta_{21}In_{(I, I + 4, A)}$ and $\Delta_{21}In_{(I, I + 6, A)}$ tell  the difference between $In$ of fragments with larger neutron-excess. The  $\Delta_{21}In_{(I, I + 4, A)}$ and $\Delta_{21}In_{(I, I + 6, A)}$ for the $^{48}$Ca/$^{40}$Ca and $^{64}$Ni/$^{58}$Ni reactions are plotted in Fig. \ref{Endif4} and Fig. \ref{Endif6}, respectively. The trends of the $\Delta_{21}In_{(I, I + 4, A)}$ and $\Delta_{21}In_{(I, I + 6, A)}$ are similar to those of $\Delta_{21}In_{(I, I + 2, A)}$. The plateau also appears in $\Delta_{21}In_{(I, I + 4, A)}$ for both of the reactions, but disappears in $\Delta_{21}In_{(I, I + 6, A)}$ for the $^{48}$Ca/$^{40}$Ca reactions. The $\Delta_{21}In_{(I, I + 4, A)}$ [$\Delta_{21}In_{(I, I + 6, A)}$] for fragments with different $I$ are consistent for the reactions. It is found that the plateau of $\Delta_{21}In_{(I, I + 4, A)}$ is almost twice that of $\Delta_{21}In_{(I, I + 2, A)}$. It is interesting if $\Delta_{21}In_{(I, I + 2, A)}$, $\Delta_{21}In_{(I, I + 4, A)}$ and $\Delta_{21}In_{(I, I + 6, A)}$ have a simple relationship. To see the systematic behavior of $\Delta_{21}In_{(I, I + m, A)}$ ($m = $ 2, 4, and 6), a scaling parameter ($S_{\Delta_{21}In}$) is defined as,
\begin{equation}\label{Sdelta21In}
S_{\Delta_{21}In} =\Delta_{21}In_{(I, I + m, A)}/(m/2), \hspace{0.5cm} m= 2, 4, 6, \cdot\cdot\cdot
\end{equation}
The results of $S_{\Delta_{21}In}$ for the $^{48}$Ca/$^{40}$Ca and $^{64}$Ni/$^{58}$Ni reactions are plotted in Fig. \ref{DDScaled}. From Fig. \ref{DDScaled}(a), it is seen that, with different $m$, $S_{\Delta_{21}In}$ for the $^{48}$Ca/$^{40}$Ca reactions is well scaled. In Fig. \ref{DDScaled}(b), similar scaling phenomenon for the $^{64}$Ni/$^{58}$Ni reactions can be found, but the range of $S_{\Delta_{21}In}$ for isobars is a little larger than that for the $^{48}$Ca/$^{40}$Ca reactions.

It is interesting to know what physics is behind the $S_{\Delta_{21}In}$ scaling phenomenon. We would like to recall the isoscaling phenomenon of fragments in HICs \cite{HShanPRL}.  The isoscaling phenomenon has been explained using many models, such as the statistical equilibrium approach \cite{AlbNCA85DRT}, the canonical ensemble theory \cite{GrandCan,Tsang07BET}, or a dubbed m-scaling in the Landau theory \cite{H-mscaling}, which all require the thermal equilibrium of system. The isoscaling parameters are related to the chemical potential difference between neutrons or protons,  the symmetry energy, or the nuclear density of the reactions \cite{HShanPRL,PMar12PRCIsob-sym-isos,PMar-IYR-sym13PRC,H-mscaling,murhodiff,MBTsPRL01iso,AMD03}. In a recent work,  $\Delta_{21}In_{(I, I + 2, A)}$ has been explained as the meaning of the isobaric yield ratio difference (IBD) probe developed in the framework of canonical ensemble theory \cite{MaPLB15}. In the canonical ensemble theory, $\sigma(I, A)$ is written as \cite{GrandCan,Tsang07BET},
\begin{equation}\label{yieldGC}
\sigma(I, A) = CA^{\tau}exp\{\beta[-F(A, I) + \mu_{n}N + \mu_{p}Z]\},
\end{equation}
where $C$ is a system-dependent constant; $\beta$ is the reverse temperature; $\mu_n$ ($\mu_p$) is the chemical potential of neutrons (protons) varying with the asymmetry of reaction system; $F(A, I)$ is the free energy of fragment. In the isobaric yield ratio, some terms depending on the asymmetry of the reaction system can be cancelled out \cite{Huang10,IBD13PRC,IBD13JPG,MaCW12PRCT,MaCW13PRCT,MaCW11PRCIYR,MaCW12EPJA}. Replacing $In(I, A)$ by $\sigma(I, A)$, the following equations can be obtained,
\begin{eqnarray}
&\Delta_{21}In_{(I, I + 2, A)} = ~\beta[\mu_{n1} - \mu_{n2}-(\mu_{p1} - \mu_{p2})], \label{GC-DIn21}\\
&\Delta_{21}In_{(I, I + 4, A)} = 2\beta[\mu_{n1} - \mu_{n2}-(\mu_{p1} - \mu_{p2})], \label{GC-DIn41}\\
&\Delta_{21}In_{(I, I + 6, A)} = 3\beta[\mu_{n1} - \mu_{n2}-(\mu_{p1} - \mu_{p2})], \label{GC-DIn61}\\
\cdot\cdot\cdot  \nonumber
\end{eqnarray}
In Ref. \cite{MaPLB15}, the relationship $\Delta_{21}In_{(I, I + 2, A)} = -\Delta(\beta\mu)_{21}$ has been illustrated, in which the IBD-$\Delta(\beta\mu)_{21}$ reflects the nuclear density or chemical potentials of neutrons (protons) for the reaction system \cite{IBD13PRC,IBD13JPG,IBD14Ca,IBDCa48EPJA,IBD15AMD,NST2015IBD}.  Eqs. (\ref{GC-DIn21}), (\ref{GC-DIn41}), and (\ref{GC-DIn61}) can be summarized as,
\begin{eqnarray}\label{GC-Scaled}
\Delta_{21}In_{(I, I + m, A)} &= (m/2)\times \beta[(\mu_{n1}- \mu_{n2})  \hspace{1cm}  \nonumber\\
&-(\mu_{p1} - \mu_{p2})], m= 2, 4, 6, \cdot\cdot\cdot 
\end{eqnarray}
$\Delta_{21}In_{(I, I + m, A)}$ thus represents the difference of $\mu_n$ and $\mu_p$ between the two reactions, which is related to the densities of neutrons and protons and nuclear symmetry energy \cite{HShanPRL,IBD14Ca,IBDCa48EPJA,NST2015IBD,murhodiff,MBTsPRL01iso}. If $\mu_n$ ($\mu_p$) or nuclear density is uniform, a perfect $S_{\Delta_{21}In}$ scaling can be expected. Then $S_{\Delta_{21}In}$ scaling reflects the degree of uniform of the nuclear matter in the reaction systems. For the $^{48}$Ca/$^{40}$Ca reactions, $S_{\Delta_{21}In}$  shows a better scaling phenomenon than that for the $^{64}$Ni/$^{58}$Ni reactions.

To further study the $S_{\Delta_{21}In}$ scaling theoretically, the antisymmetric quantum dynamics (AMD) model \cite{AMD96,AMD99,AMD03,AMD04} plus the sequential decay code GEMINI \cite{gemini} are used to simulate the 140$A$ MeV $^{58, 64}$Ni + $^9$Be reactions. The standard Gogny (Gogny-g0) interaction \cite{g0-pot} is used in the AMD simulations. The fragments are formed with a coalescence radius R$_c =$ 5 fm in the phase space at $t =$ 500 fm/c \cite{IBD13PRC,IBD15AMD,Ma15CPL}. Since no global equilibrium can be obtained in transport models, it is not safe to use the thermodynamics probes directly in the AMD simulations. But the information uncertainty probe does not require an equilibrium system, which still can be used in the reactions simulated by the AMD model. The $S_{\Delta_{21}In}$ for the simulated $^{64}$Ni/$^{58}$Ni reactions are plotted in Fig. \ref{DDScaledAMD}, which has a similar distribution as that for the measured $^{64}$Ni/$^{58}$Ni reactions. For the AMD + GEMINI simulated reactions, the $S_{\Delta_{21}In}$ shows a better scaling phenomenon compared to that for the AMD + GEMINI simulated reactions since there is a larger fluctuation in the AMD results. It is known that the decay process will influence the results \cite{Huang10}. Meanwhile, the plateaus of $S_{\Delta_{21}In}$ in the measured reactions is also reproduced in the simulated reactions, which are illustrated in the shadowed areas. It can be concluded that the AMD (+ GEMINI) simulations well reproduce the scaling phenomenon in $S_{\Delta_{21}In}$.

It should be stressed that the concept of information uncertainty, which does not require an equilibrium system, can be used both in the thermodynamic models and the transport models. Having a similar form to the thermodynamic probes, the information uncertainty probe has the advantage to extract information of evolving reaction and can be directly compared to the final state of the reaction, which corresponds to simulation of the thermodynamic models. The information uncertainty probe bridges the results of transport models and thermodynamic models by avoiding the abrupt use of the thermodynamic probes, which sheds new light on the researches in HICs.

\section{summary}
To summarize, the Shannon information-entropy uncertainty is used to study the products in HICs. By defining the difference of information uncertainty between two reactions, a scaling phenomenon of $S_{\Delta_{21}In}$ is found in the fragments of the $^{48}$Ca/$^{40}$Ca and $^{64}$Ni/$^{58}$Ni reactions. The $S_{\Delta_{21}In}$ scaling is explained by the concept of the canonical ensemble theory, which indicates that $S_{\Delta_{21}In}$ reflects the properties of nuclear matter in HICs. In addition, the $S_{\Delta_{21}In}$ scaling is also proven in the simulation of the 140$A$ MeV $^{58, 64}$Ni + $^9$Be reactions by the AMD (+ GEMINI) models. The probe for nuclear matter based on information uncertainty, which has a similar form as the thermodynamic models, can be used in the non-equilibrium system in the dynamical process, and connect the results between transport models and thermodynamics models.

\section*{acknowledgments}
\textit{This work is supported by the Program for Science \& Technology Innovation Talents in Universities of Henan Province (13HASTIT046), and the National Natural Science Foundation of China (Nos. 11421505, 11035009, and 11205079), the Knowledge Innovation Project of the Chinese Academy of Sciences (No. KJCX2-EW-N01).
C.-W. Ma thanks Prof. R. Wada and A. Ono for providing us the AMD code, and thanks Dr. M. Huang for the help in the AMD simulations. C.-W. Ma also thanks the kindly hospitality from Prof. Natowitz during Ma's visiting to Texas A\&M University.
}



\begin{thebibliography}{5}
%
%
\bibitem{Shannon}
C. E. Shannon,
Bell System Technical Journal {\bf 27}, 379 (1948).

\bibitem{Jaynes}
E. T. Jaynes, Phys. Rev. {\bf 106}, 620 (1957).
\bibitem{EntrHC}
P. Brogueira, J. Dias de Deus, and I. P. da Silva,
Phys. Rev. D \textbf{53}, 5283 (1996);\\
Z. Cao and R. C. Hwa, Phys. Rev. D \textbf{53}, 6608 (1996).
\bibitem{YGMaZipfPRL99}
Y. G. Ma, Phys. Rev. Lett. \textbf{83}, 3617 (1999).
\bibitem{InfEnt02}
P. Balenzuela and C. O. Dorso, Phys. Rev. C \textbf{65}, 057602 (2002).
\bibitem{InfEnt04}
A. Barra\~{n}\'{o}n, J. E. Roa, and J. A. L\'{o}pez,  Phys. Rev. C \textbf{69}, 014601 (2004).
\bibitem{MaPLB15}
C. W. Ma, H. L. Wei, S. S. Wang, Y.G. Ma, R. Wada, Y. L. Zhang,
Phys. Lett. B \textbf{742}, 19 (2015).
\bibitem{MaCTP14}
C. W. Ma, H. L. Wei, Commun. Theor. Phys. \textbf{62}, 717 (2014).
\bibitem{Mocko06}
M. Mocko, M. B. Tsang, L. Andronenko
{\it et al.},
Phys. Rev. C \textbf{74}, 054612 (2006).
\bibitem{HShanPRL}
H. S. Xu \textit{et al.}, Phys. Rev. Lett. \textbf{85}, 716 (2000).
\bibitem{AlbNCA85DRT}
S. Albergo, S. Costa, E. Costanzo, and A. Rubbino,
Nuovo Cimento A {\bf 89}, 1 (1985).
\bibitem{GrandCan} 
C. B. Das, S. Das Gupta, X. D. Liu, and M. B. Tsang,
Phys. Rev. C \textbf{64}, 044608 (2001).
\bibitem{Tsang07BET} 
M. B. Tsang, W. G. Lynch, W. A. Friedman
\textit{et al.},
Phys. Rev. C \textbf{76}, 041302(R) (2007).

\bibitem{H-mscaling}
M. Huang, Z. Chen, S. Kowalski, R. Wada, T. Keutgen
{\it et al.},
Nucl. Phys. A \textbf{847}, 233 (2011). 
\bibitem{PMar12PRCIsob-sym-isos} 
P. Marini, A. Bonasera, A. McIntosh, R. Tripathi, S. Galanopoulos
{\it et al.},
Phys. Rev. C \textbf{85}, 034617 (2012).
\bibitem{PMar-IYR-sym13PRC} 
P. Marini, A. Bonasera, G. A. Souliotis, P. Cammarata, S. Wuenschel
{\it et al.},
Phys. Rev. C \textbf{87}, 024603 (2013).



\bibitem{murhodiff}
E. Geraci, M. Bruno, M. D'Agostino
\textit{et al.},
Nucl. Phys. A \textbf{732}, 173 (2004).
\bibitem{MBTsPRL01iso}
M. B. Tsang, W. A. Friedman, C. K. Gelbke, W. G. Lynch,
G. Verde, and H. Xu,
Phys. Rev. Lett. \textbf{86}, 5023 (2001).
\bibitem{AMD03}
A. Ono, P. Danielewicz, W. A. Friedman, W. G. Lynch, and M. B. Tsang,
Phys. Rev. C \textbf{68}, 051601(R) (2003).

\bibitem{Huang10} 
M. Huang, Z. Chen, S. Kowalski
{\it et al.},
Phys. Rev. C \textbf{81}, 044620 (2010).

\bibitem{MaCW12PRCT} 
C. W. Ma, J. Pu, Y. G. Ma, R. Wada, S. S. Wang,
Phys. Rev. C \textbf{86}, 054611 (2012).
\bibitem{MaCW13PRCT}
C. W. Ma, X. L. Zhao, J. Pu 
{\it et al.},
Phys. Rev. C \textbf{88}, 014609 (2013).

\bibitem{MaCW11PRCIYR}
C. W. Ma, F. Wang, Y. G. Ma and C. Jin,
Phys. Rev. C \textbf{83}, 064620 (2011).
\bibitem{MaCW12EPJA}
C.-W. Ma, J. Pu, H.-L. Wei
{\it et al.},
Eur. Phys. J. A \textbf{48}, 78 (2012). 

\bibitem{IBD13PRC}
C. W. Ma, S. S. Wang, Y. L. Zhang, H. L. Wei, Phys. Rev. C \textbf{87}, 034618 (2013).
\bibitem{IBD13JPG}
C.W. Ma, S. S. Wang, Y. L. Zhang, H. L. Wei, J. Phys. G: Nucl. Part. Phys. \textbf{40}, 125106 (2013).
\bibitem{IBDCa48EPJA}
C. W. Ma, X. M. Bai, J. Yu, H. L. Wei,
Eur. Phys. J. A \textbf{50}, 139 (2014).

\bibitem{IBD14Ca}
C.W. Ma, J. Yu, X. M. Bai, Y. L. Zhang, H. L. Wei, S. S. Wang,
Phys. Rev. C \textbf{89}, 057602 (2014).
\bibitem{IBD15AMD}
C.-Y. Qiao, H. -L. Wei, C.-W. Ma, Y.-L. Zhang, S.-S. Wang
\textit{et al.},
Phys. Rev. C \textbf{92}, 014602 (2015).
%
\bibitem{NST2015IBD}
M. Yu, K.-J. Duan, S.-S. Wang 
\textit{et al.},
Nucl. Sci. Tech. \textbf{26}, S20503 (2015).
\bibitem{AMD96}
A. Ono and H. Horiuchi, Phys. Rev. C \textbf{53}, 2958 (1996).
\bibitem{AMD99}
A. Ono, Phys. Rev. C \textbf{59}, 853 (1999).
\bibitem{AMD04}
A. Ono, P. Danielewicz, W. A. Friedman, W. G. Lynch, and M. B. Tsang,
Phys. Rev. C \textbf{70}, 041604(R) (2004).
\bibitem{gemini}
R. J. Charity \textit{et al.},
Nucl. Phys. A \textbf{483}, 371 (1988).
\bibitem{g0-pot}
J. Decharg\'{e} and D. Gogny,
Phys. Rev. C \textbf{21}, 1568 (1980).
\bibitem{Ma15CPL}
C.-W. Ma, Y.-L. Zhang, S.-S. Wang, and C.-Y. Qiao,
Chin. Phys. Lett. \textbf{32}, 072501 (2015).


\end{thebibliography}
\end{document}